\documentclass[fleqn,twoside]{article}
\usepackage{espcrc2}

\usepackage{graphicx}
\usepackage[figuresright]{rotating}

\newcommand{\AmS}{{\protect\the\textfont2
  A\kern-.1667em\lower.5ex\hbox{M}\kern-.125emS}}

\title{Sub-horizon Perturbation Behavior in  Extended Quintessence}

\author{F. Perrotta \address[OAP]{INAF-Osservatorio Astronomico di Padova, 
Vicolo dell'Osservatorio 5, 35122 Padova, Italy, and \\
Lawrence  Berkeley  National Laboratory, 1 Cyclotron Road, M/S 
50-205, Berkeley, CA 94720},
        C. Baccigalupi \address{SISSA-ISAS, Via Beirut 1-4, 34014 Trieste, 
        Italy, and \\
Lawrence  Berkeley  National Laboratory, 1 Cyclotron Road, M/S
50-205, Berkeley, CA 94720} }

\begin{document}

\begin{abstract}
In the general context of scalar-tensor theories, we consider a model in 
which a scalar field coupled to the Ricci scalar in the gravitational 
sector of the Lagrangian, is also playing the role of an ``Extended 
Quintessence''  field, dominating the energy content of the Universe  at 
the present time. In this framework, we study the linear evolution 
of the perturbations in the Quintessence energy density, showing that a 
new phenomenon, named here  ``gravitational dragging'', can enhance the 
scalar field density perturbations as much as they reach the non-linear 
regime. The possibility of dark energy clumps formation is thus discussed.  
\vspace{1pc}
\end{abstract}

\maketitle

\section{The growth of quintessence linear perturbations.}
\begin{figure}[htb]
\hspace{-26pt}
\includegraphics[height=100mm]{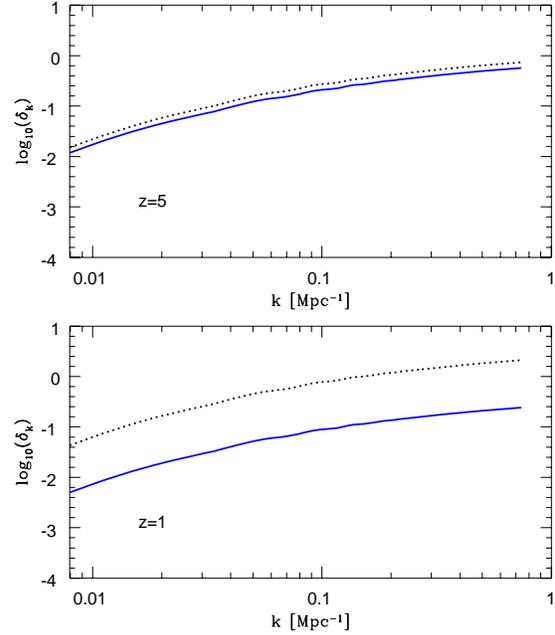}
\caption{Power spectrum of linear perturbation in CDM component (dotted 
line) and EQ component (solid) at redhifts z=5 (upper panel) and z=1 (lower 
panel).} 
\label{powerspectrum}
\end{figure}

The recent growing evidences for a vacuum energy component in the Universe 
\cite{PERL,RIESS} have lead to the development of a number of models, most 
of which tend to alleviate the theoretical difficulties of the well known 
``cosmological constant problem'' (see \cite{Car} for a review).  All these 
models are replacing the cosmological constant with a ``dynamical vacuum energy'' 
provided by an evolving scalar field, often named ``Quintessence''.  
This field should provide the present cosmic accelerating phase through its 
negative equation of state. 
Such ``dark energy'' component has also been modeled in the general 
framework of scalar-tensor theories of Gravity: in these ``Extended 
Quintessence''  (EQ) scenarios \cite{EQ}-\cite{CHIBACOS}, the Quintessence 
scalar field is directly  coupled to the Ricci scalar $R$ in the Lagrangian 
of the theory: the  canonical Ricci scalar term $\kappa ^{-1} R$, with 
$\kappa= 8 \pi G_*$, is replaced by $F(\phi) R$, where $F(\phi)$ is a 
function of the quintessence  field $\phi$. Here, $G_*$ is a ``bare'' 
gravitational constant, generally different from the Newton's constant G 
as it is measured by Cavendish-type experiments \cite{RU,EF}. 
In EQ models, a scalar field has thus a double role: 
at any epoch, its value determines 
the effective gravitational constant, and the contribution in dark energy 
density. 
Of course, such a coupling is not arbitrary: it is instead 
subject to a number of constraints, mainly arising from the bounds on the 
time variations of the constants of nature, which must be taken into 
account.  EQ scenarios have been analyzed from different 
points of view, and there are robust predictions about their possible 
imprints on the power spectrum of CMB anisotropies \cite{EQ,TEQ}. 
However, the important role of Dark Energy in modern cosmology opens the 
question of its relation with the other main dark component currently under 
investigation, the Dark Matter. 
One major point with Quintessence scenarios is whether a Dark Energy component 
could affect, in some way, the formation of Dark Matter clumps. In 
particular, one fundamental question is: can the
Dark Energy itself form clumps on certain scales? If we restrict ourselves 
to minimally-coupled scalar fields, we have to deal with a very 
relativistic component: indeed, the sound speed of such component turns 
out to be identical to that of radiation \cite{GDM}. As a 
consequence, an ``ordinary'' (i.e., minimally coupled) Quintessence 
component becomes homogeneous on scales  smaller than the horizon, as its 
energy density perturbations are rapidly damped out. For this reason, 
clumps of ``ordinary'' Quintessence cannot form on such scales. 
In EQ scenarios, however, the situation can dramaticaly
change, since the physical properties of an EQ are deeply
modified with respect to those of an ordinary Quintessence field 
\cite{DEC}, as is the case for its effective sound speed;
one natural question is whether dark energy structures could form in 
these scenarios, starting from initial small overdensities in the linear 
regime. In order to give an answer to this question, we start from a 
definition of the stress-energy tensor $T_{\mu \nu} (\phi)$ of the 
Quintessence field, which includes a term of purely gravitational origin, in 
addition to the standard contributions from minimal coupling and non-minimal 
coupling \cite{Faraoni,Torres}: 
$$
T_{\mu \nu}[\phi] =T_{\mu \nu}^{mc}[\phi]+  T_{\mu \nu}^{nmc}[\phi] +
T_{\mu \nu}^{grav}[\phi] \ \ , 
$$
with 
$$
T_{\mu \nu}^{grav}[\phi] =\left( {1\over \kappa}-F\right) G_{\mu \nu}\ .
$$
The so-defined  stress-energy tensor of the Quintessence field is a 
conserved  quantity, as a consequence of the contracted Bianchi identities; 
drawing the scalar field energy density $\rho_{\phi}$ from this tensor, 
it will satisfy the continuity equation. Most importantly, also in 
$\rho_{\phi}$ we can  distinguish the gravitational contribution from the 
others, analogously to the separation performed into $T_{\mu \nu}(\phi)$: 
$$
\rho_{\phi}={{\dot{\phi}}^2 \over 2 a^2}+V(\phi)-
{3\over a^{2}}{\cal H}\dot{F}+3{{\cal H}^{2}\over a^{2}}
\left( {1\over\kappa}-F \right)\ . 
$$
In the last expression, ${\cal{H}}$ is the conformal expansion rate, and 
$V(\phi)$ is the scalar field potential. 
Let us focus on the last term in the rhs of this equation: noticeably, when 
$F(\phi) \neq \kappa^{-1}$, the gravitational contribution to $\rho_{\phi}$ 
switches on,  feeding $\rho_{\phi}$ with a term proportional to the total 
energy density of the universe via ${\cal{H}}^2$. In practice, the 
Quintessence energy density is dragged by the energy density 
of the dominant component at any epoch: this effect, which we call 
``gravitational dragging'', is also active on the time evolution of the 
perturbations.  Indeed, we could perform an analogous division between the 
various contributions to the scalar field density contrast $\delta_{\phi} 
\equiv \delta \rho_{\phi} / \rho_{\phi} $, finding 
$$
\delta_{\phi}=\delta_{\phi}^{mc}+\delta_{\phi}^{nmc}+\delta_{\phi}^{grav}\ \ 
, 
$$
being the gravitational contribution \cite{DEC}
$$
\delta_{\phi}^{grav}={-6\delta F{\cal H}^{2}-
2a^{2}(1/\kappa -F)\delta G_{0}^{0}
\over
\dot{\phi}^{2}+2a^{2}V-6{\cal H}\dot{F}-2a^{2}
(1/\kappa -F)G_{0}^{0}}\ .
$$  
Looking at this expression, it is evident that the Quintessence field 
density contrast can be non-vanishing  even when the Quintessence field 
is homogeneous by itself, provided $F \neq \kappa^{-1}$; in this case, 
indeed, $\delta \rho_{\phi}$ is sourced by the $\delta G^0_0$ term, 
containing the total cosmic fluid energy density perturbations, which are 
generally non vanishing.  While the scalar field perturbations 
$\delta \phi$ contribute to $\delta_{\phi}^{mc}$ and $\delta_{\phi}^{nmc}$ 
only, the Ricci scalar perturbations are responsible for the gravitational 
contribution to $\delta \rho_{\phi}$; the latter can dominate the 
dynamics of  $\delta \rho_{\phi}$ whenever $\delta \phi, \delta \dot{\phi}$ 
are negligible and $F \neq \kappa^{-1}$.   
In other words, the  gravitational dragging mechanism  forces  the scalar 
field perturbations to behave as the density perturbations of the dominant 
cosmic fluid component: noticeably, in the matter dominated era, an EQ
field can even behave as pressureless dark matter, thanks to 
the explicit coupling netween $\phi$ and $R$. 
The Quintessence field overdensities receive an injection of power which may 
potentially enhance them up to the exit from the linear regime, following 
the matter behaviour. In fig. \ref{powerspectrum} we plot the logarithmic 
power of density fluctuations, defined in the Fourier space as $\delta^2_k
=4 \pi k^3 \delta^2$, both for CDM perturbations and for EQ perturbations, 
in a specific example with $F(\phi)= \kappa^{-1} + \zeta (\phi^2)$
(where $\zeta$ is an observationally constrained coupling constant). Note 
that, at the present time,  the quantity $\kappa^{-1}-F$ reduces to 
$\zeta \phi_0^2$ (where $\phi_0$ is the field value at $z=0$), 
which is generally a non-vanishing term.  
The plots in figure \ref{powerspectrum} refers to redshifts $z=5$ and 
$z=1$; because of the gravitational dragging, linear 
perturbations in the Quintessence energy density are close to those in 
the CDM component up to redshifts relevant for structure formation, while 
they appear to decrease at $z \sim 1$, when the quintessence field  
comes to dominate the expansion and the term  $(\kappa^{-1}-F(\phi))G^0_0$ 
is no more the dominant term in the denominator of $\delta_{\phi}^{grav}$. 
However, scales on which perturbations reached the 
non-linear regime at some redshift are no more adequately described by the 
linear approach. 
At that point, it is quite obvious that a non-linear 
approach is required to further investigate the Quintessence perturbation 
evolution. \\
The same result can be obtained with a different approach, based on the 
properties of the ``effective sound speed'' $c_{eff}^2$ of a fluid 
\cite{GDM}: 
such a  quantity, defined as the sound speed in the fluid rest frame, 
encloses both  the adiabatic and entropic contributions to the sound speed, 
and it well characterizes the total pressure response to the gravitational 
collapse. 
While it is equal to the speed of light for minimally-coupled scalar fields, 
it turns out to be a scalar field dependend quantity in EQ models; in the gravitational dragging regime, the effective 
sound speed is inversely proportional to $\delta \rho_{\phi}$, which can be 
at the level of perturbations in the CDM component, resulting in low values 
of $c_{eff}^2$. Also the sign of $c_{eff}^2$ may be reversed, depending on 
the 
sign of the coupling constant, strongly modifying the response of pressure 
to the variation in the energy density, and radically changing the 
gravitational collapse scenario for the Quintessence component.    

The fact that the Quintessence sound speed gets strongly modified by the 
coupling to the Ricci scalar, a result which is analogous to the 
gravitational dragging mechanism described above, allows for the possibility 
of Quintessence clumps; the characteristic  scales for this process have 
not been investigated yet,  and the further evolution of Quintessence energy 
density  perturbations in the non-linear regime deserve a future 
investigation. 

\section{Possible implications of Quintessence clumpiness.}
The intriguing possibility of the gravitational dragging of matter 
perturbations on Quintessence perturbations is one of the possible grounds 
where to probe the relation between Dark Matter (DM) and Dark Energy (DE). In 
recent  years, the possibility of DE clumps formation, the stability  
of spherically symmetric scalar field structures and their relation to 
the DM component in galaxies, have received increasing attention 
\cite{DEC,WETTECLUMPS}.
In the  scenario here proposed, such relation can only be of gravitational 
type, and may possibly be detected on the galactic scales.  
The question of how the DM is distributed inside dark halos and   
cluster is still matter of debate. For example, it is still unknown whether 
the dark halos density profile is universal, or it depends on properties of 
the galaxy itself.   Kinematic properties, such as rotation curves of disk 
galaxies,  are an excellent  probe for DM on galactic scales; it has been 
shown that  independent   high-resolution optical rotation curves 
observations, can probe the halo mass distribution and resolve their 
structure. An important contribution
has been obtained \cite{BS} by the analysis of low-luminosity late-type 
spirals; the surprising
result of this analysis has been that the halo velocity profiles increase  
linearly with the radius, at least up to the stellar disk edge (the  
``visible galaxy").  This implies that, over all the stellar disk, the 
dark halo density is about a constant; similar properties were previously 
found
also for dwarf and low surface brightness galaxies, which are both DM 
objects \cite{Moore,FP}.
The presence of a constant density central core is in contrast with the 
CDM scenario, which predicts a steeply cusped density
distribution in the same region. The presence of a 
central singularity in the density profile of DM has to be   
excluded also as a result of the analysis of gravitational lensing on the  
cluster scales.  This disparity between the model and the data on the
galaxy rotation curves requires the introduction of different fields
and/or particles, including the DE as a candidate. Indeed, this   
disparity follows directly from the nature of collisionless particles, 
which have no physically associated length scale, with the result that 
DM halos should have zero effective core radii \cite{Moore}. In order to 
generate cores in the DM distribution, additional physical 
properties are required, such as could be provided by Hot Dark Matter (in 
this case, the  central density may be naturally limited by phase space 
constraints) or an universe dominated by dissipative baryonic matter. 
Even though these models 
have to be excluded because they fail in accounting for the observed 
Large Scale Structure, they are both able to provide a 
mechanism that limits the central density of the CDM particles, 
introducing a characteristic scale. In this context, it is extremely 
interesting the  question of how, and whether, the DE has a key 
role into the structure of the DM halos. If this is the case, we 
expect that  this role is based on the purely gravitational interaction 
between DM and DE, which may imply a redistribution 
of DM in  the halos, particularly in the central core.

The gravitational dragging  mechanism described above opens new perspectives 
on the relation between DM and DE in 
collapsed structures, from galactic scales to the scales of galaxy 
clusters.  In particular,  predictions for the galactic rotation curves in 
EQ models should be drawn and compared to observations. 
A non-linear approach to the perturbation growth in gravitational dragging 
regime will allow to check for the possibility of constant density central 
cores in  galaxies, and for determining the characteristic scales of 
eventual Quintessence clumps.

\end{document}